\documentstyle[preprint,revtex]{aps}
\tightenlines
\begin{document}
\draft
\begin{title}
Length-scale competition in the
damped sine-Gordon chain\\
with spatio-temporal periodic driving
\end{title}
\author{David ~Cai$^{(1)}$, A. R.~ Bishop$^{(1)}$,
and Angel ~S\'{a}nchez$^{(1),(2)}$}
\begin{instit}
$^{(1)}$Theoretical Division and
Center for Nonlinear Studies, \\Los Alamos National Laboratory,\\ Los Alamos,
New Mexico 87545, U.S.A.,\\
    \\
$^{(2)}$Escuela Polit\'{e}cnica Superior
\\Universidad Carlos III de Madrid
\\Avda. del Mediterr\'{a}neo 20
\\E-28913 Legan\'{e}s, Madrid, Spain
\end{instit}
\begin{abstract}
It is shown that there are two different regimes for the
damped sine-Gordon chain driven by the spatio-temporal periodic force
$\Gamma sin(\omega t - k_{n} x)$ with a flat initial condition. For
$\omega > k_{n}$, the system first bifurcates at a critical
$\Gamma_{c}(n)$ to a translating {\em 2-breather} excitation from a
state locked to the driver. For $\omega < k_{n}$, the excitations of
the system are the locked states with the phase velocity $\omega/k_{n}$
in all the region of $\Gamma$ studied.
In the first regime, the frequency of the
breathers is controlled by $\omega$, and the velocity of the
breathers, controlled by $k_{n}$, is shown to be the group velocity
determined from the linear dispersion relation for the sine-Gordon
equation. A linear stability analysis reveals that, in addition to two
competing length-scales, namely, the width of the breathers and the spatial
period of the driving, there is one more length-scale  which
plays an important role in controlling the dynamics of the system at
small driving.
 In the second regime the length-scale $k_{n}$ controls
the excitation. The above picture is further corroborated by
numerical nonlinear spectral analysis.  An energy balance estimate is
also presented and shown to predict the critical value of
$\Gamma$ in good agreement with the numerics.

\end{abstract}

\begin{narrowtext}
\narrowtext

\noindent
I. INTRODUCTION
\vspace{20pt}

Recently it has been shown that  length-scale competition
is crucial for understanding dynamics of  both  one-dimensional
nonlinear Schroedinger (NLS) and one-dimensional
sine-Gordon (SG) systems driven by a static, spatially periodic
parametric potential\cite{Scharf,Sanchez0,Scharf1}. The picture that has
emerged is that the length-scale competition between the width of a
breather and the spatial period of the external potential controls
the particle-like coherence of the breather.
The work of Refs
\cite{Scharf,Sanchez0,Scharf1}
 has also established criteria for the validity of
collective coordinates descriptions for the excitations in the systems.
In the work of Refs\cite{Ariyasu1,Ariyasu2} it has been
 demonstrated that a damped
SG chain with additive dc-driving
 also exhibits length-scale competition.

In this article  we study the periodic SG system additively driven by a {\em
spatio-temporal} periodic potential with a small damping strength.
We anticipate that two lengths are intrinsic to this system and can
be separately controlled, one being the width of a breather excited by
 the driving, and the other the spatial
period of the driving field. The creation of a  breather mode
can be expected  from the
extensive work on the one-dimensional damped SG system additively
driven by a temporally periodic external force
\cite{Bishop1,Mazor1,Mazor2,Bishop2,Bishop3,Bishop4,Terrones,Gronbech-Jensen1,Gronbech-Jensen2}.
 This work has shown that the system exhibits
transitions from oscillating spatially uniform states locked to the
periodic driving force, to solitonic breather excitations, to more
spatially irregular, less coherent states which are in a low
dimensional chaotic regime described in terms of a small number of
interacting collective modes. The frequency
of the coherent breather mode, excited at the first bifurcation with
increasing driving strength, is locked to the frequency of the
driving. This frequency-locking mechanism enables us to {\em dynamically}
control the width of a breather in our system, {\it i.e.} by
controlling the temporal frequency of the driving.

In light of the works mentioned above, one might expect that: (i) our
system will lock to the external driving; (ii) above a critical strength
there will be additional breather-type excitations whose time scale
is controlled by the driving frequency, and (iii) there
will be a spatial length-scale competition between the width
of a breather and the spatial period of the driving.
 Our investigation supports this scenario and
reveals an additional spatial length-scale that plays a significant role in
the competitions between length scales in the system, namely,
the wavelength of the lowest spatial modulational mode.

Direct numerical simulation has provided much insight for
the behavior of our system. Combined with linear stability analysis,
energy balance estimates \cite{Mazor2,Lomdahl},
 and nonlinear spectral analysis \cite{Overman,Forest,Flesch}, we have been
able to achieve a broad understanding.

The organization of this article is as follows. In section II we define
the model and summarize the main numerical results;  In section III
 we investigate
the linear stability of the locked periodic traveling wavetrain
 and show the origin of one of the spatially
unstable modes; In section  IV, by a simple energy balance argument,
 we provide a further understanding of
 the transition from the
periodic traveling wavetrains to a breather-mode excitation; In section V
we describe nonlinear spectral analysis results for our system.

\vspace{24pt}
\noindent
II. THE MODEL AND THE NUMERICAL RESULTS
\vspace{20pt}

The governing equation of our system is:
\begin{equation}
U_{tt} - U_{xx} + \sin U = - \alpha U_{t} + \Gamma \sin(\omega t - k_{n} x )
\label{eq:SG}
\end{equation}
\[\exp (iU(x,t)) = \exp (i U(x+L,t)) \]
\[U(x,t=0) = U_{t}(x,t=0) =0, \]
where $L$ is the total length of the system, $\alpha$ is the damping
coefficient, $\Gamma$ is the driving strength, and
$k_{n} = 2 \pi n/L$. We chose
$\omega = 0.9$, and $\alpha = 0.1$ and studied the cases for $1 \leq n \leq 7$
and $L = 24$. With this choice of $\omega$, we are
in the low-amplitude, near-NLS, regime of SG.
Occasionally we will compare the results for $L=24$ with those
for $L=36 $ to give a more complete picture.

The method we used to numerically integrate the equations is
based on the Strauss-V\'{a}zquez scheme (\cite{Vazquez}; See S\'{a}nchez
{\it et al}. \cite{Sanchez} for details and references on its application
to perturbed nonlinear Klein-Gordon problems).
Fig.~\ref{f1} summarizes the main numerical observations for $L= 24$.
 There is a major separation between two regimes, Regime
I: $n \leq 3$, {\it i.e.} $k_{n} < \omega$,
and Regime II: $n >3$, {\it i.e.} $k_{n} > \omega$.
For Regime II, if we seek an ansatz of the form $U(\omega t - k_{n} x)$,
Eq(~\ref{eq:SG}) is reduced to a nonlinear oscillator with
an {\em anti}-damping
coefficient, {\it i.e.}
\[u_{\zeta\zeta} + \sin u = \alpha^{\prime} u_{\zeta} + \Gamma\sin(W\zeta), \]
where $W = \sqrt{k_{n}^{2} - \omega^{2}}$, $\alpha^{\prime} = \alpha\omega/W$,
$W\zeta = \omega t - k_{n} x - \pi$, $u = U - \pi$.
A corollary of the
linear stability analysis outlined below is that there is no linear
stability for a wavetrain of the form $U_{0}\cos(\omega t - k_{n} x)$ in this
Regime II.
Numerically
we observed only stable periodic wavetrains locked to
the external driving, with the phase velocity $\omega/k_{n}$. These
wavetrains can be viewed as breather-trains locked to
the driving.
This indicates that in this regime the $k_{n}$ mode controls the dynamics of
 the system and $k_{n}^{-1}$ is the dominant length-scale.

 Regime I exhibits far more complicated phenomena. In the
following discussion we will confine ourselves to this regime unless
otherwise specified.
As is well known, the pure SG equation,
\[U_{tt} - U_{xx} + \sin U = 0,\]
exhibits modulational instability, {\it i.e.} a periodic wavetrain of the
functional form $U=U(\omega t-k x)$ is unstable
({\it e.g.}
\cite{Whitham}). With the additional
 spatio-temporal periodic driving in Eq.(~\ref{eq:SG}),
the system
displays rather different behaviors. As for the case of the SG system
driven by a temporally periodic force (see {\it e.g.}\cite{Bishop1,Overman}),
 $n=0$ in our setting,
for small $\Gamma$ below a certain critical value $\Gamma_{c}(n)$, our system
for $ n> 0$ forms a periodic traveling wave locked to the
driving from the flat initial background, {\it i.e.} of the form
$U(\omega t - k_{n} x)$.  With increasing $\Gamma$
the system exhibits a sharp transition at $\Gamma_{c}(n)$ to a new coherent
state in which the breather excitation
 is observed and
confirmed by the  nonlinear spectral analysis.  The frequency of
the breathers is controlled by the temporal part of the external driving.
Details of the nonlinear spectral
spectral analysis will be presented in section V.

\vspace{24pt}
\noindent
III. LINEAR STABILITY ANALYSIS
\vspace{20pt}

In the following, we present a linear stability analysis and show
the origin of one of the spatially unstable modes.
 First, we seek a solution of the form
\(U = f(\omega t -
k_{n} x + \theta)\) for the linearized version of Eq(~\ref{eq:SG}) and find
\begin{eqnarray}
f  = U_{0}\cos ( \omega t - k_{n} x - \theta ) \label{eqf} \\
U_{0} = \frac{\Gamma}{\sqrt{ (k_{n}^{2} - \omega^{2} +1)^{2} + (\alpha
\omega)^{2}}} \nonumber \\
\theta = - \arctan(\frac{ k_{n}^{2} - \omega^{2} +1} {\alpha \omega}).
\end{eqnarray}
Now letting
\[U = f + \eta,\]
the linear stability is governed by
\begin{equation}
\eta_{tt} - \eta_{xx} + \cos(U_{0} \cos( \omega t- k_{n} x - \theta)) \eta +
\alpha \eta_{t} = 0. \label{LS}
\end{equation}
Introducing $z =\omega t - k_{n} x  - \theta$, we have from Eq(~\ref{LS})
\begin{equation}
- \eta_{xx} + 2 k_{n} \eta_{xz} + (\omega^{2} - k_{n}^{2}) \eta_{zz} + \alpha
\omega \eta_{z}+\cos [ U_{0} \cos(z)] \eta =0.
\end{equation}
If we are concerned only with the instability in the amplitude of
$f(\omega t -k_{n} x + \theta)$, {\it i.e.} the amplitude instability
 of the  uniform wavetrain,
 we may assume $\eta(z)$ is independent of
$x$.
Hence the omission of $\eta_{xz}$.
Alternatively, for large $L$, we may neglect the term
$2 k_{n} \eta_{xz}$.
Separating variables, we have
\begin{eqnarray}
(\omega^{2} - k_{n}^{2}) Z_{zz} + \omega \alpha Z_{z} + [K^2+\cos(U_{0}\cos
z)]Z = 0 \label{eq:Z} \\
X^{\prime\prime} + K^{2} X = 0, \label{eq:X}
\end{eqnarray}
where
$\eta = XZ$. For the amplitude instability of the uniform wavetrain,
 we have $K \equiv 0$ in Eq(~\ref{eq:X}) and
we will retain  the
$k_{n}^2$ term without making the assumption that $L$ is large.
Of course, in the large $L$ approximation,
the $k_{n}^2$ term in Eq(~\ref{eq:Z}) should be
omitted.
Now letting
\begin{eqnarray}
Z & = &  \exp[- \frac{\omega \alpha z}{2(\omega^{2} - k_{n}^{2})} ]Y,
\label{eq:expo} \\
y & = & 2z + \pi \nonumber,
\end{eqnarray}
and assuming small $U_{0}$, Eq(~\ref{eq:Z}) becomes
\begin{equation}
Y^{\prime\prime} + [\delta + \epsilon \cos y] Y=0. \label{eq:mathieu}
\end{equation}
This is the Mathieu equation with
\begin{eqnarray}
\delta & = & \frac{1}{(\omega^{2} -k_{n}^{2})} [\frac{1}{4} - \frac{
U_{0}^{2}}{16} - \frac{(\alpha\omega)^{2}}{16(\omega^{2}-k_{n}^{2})} +\frac{
K^{2}}{4}] \\
\epsilon & = & \frac{U_{0}^{2}}{16(\omega^{2}-k_{n}^{2})}.
\end{eqnarray}

The parameter space $(\delta, \epsilon)$
of the Mathieu equation (~\ref{eq:mathieu}) is covered by bands
of stable  and unstable regions.
Notice from Eq(~\ref{eq:expo}) that $\eta$ is unbounded
if $k_{n} > \omega$, since $Y(y)$
 has
either an asymptotically unbounded solution or a periodic one.
It follows that there is no linear amplitude stability for regime II.
Our system is near the
critical point \(\delta = 1/4\) for small $\epsilon$ and low $K$.
 By increasing $U_{0}$, which is proportional to $\Gamma$,
 the parameters of Eq(~\ref{eq:mathieu})
 move across the
boundary \(\delta = 1/4 + \epsilon/2 + ... \) from the stable to the
unstable region. We treat $\epsilon$ as the expansion
parameter and define $\delta_{1}$ by
\[\delta = 1/4 +\epsilon \delta_{1}.\]
{}From the general theory of the Mathieu functions \cite{Kevorkian,McLachlan},
 we have the exponential factor
\[\exp (\epsilon y\sqrt{ 1/4 - \delta_{1}^{2} })\]
 for $Y(y)$
in the asymptotic solution. Thus the boundary between the stable and the
unstable regions for Eq(~\ref{eq:Z}) is determined by
\begin{equation}
2\epsilon \sqrt{1/4-\delta_{1}^{2}} =
 \frac{\alpha\omega}{2(\omega^{2} -k_{n}^{2})}. \label{eq:critical}
\end{equation}

{}From Eq.(~\ref{eq:critical})
 we have $\Gamma_{c}(1)=0.24, \Gamma_{c}(2)=0.53$, and
$\Gamma_{c}(3)=1.19$: See Fig.~\ref{f1} and Table~\ref{T1}.
In determining $\Gamma_{c}$ we have
set $K=0$, {\it i.e.} there is no spatial modulation on the amplitude.
$\Gamma_{c}$  calculated in this way is the critical value for the onset
of the amplitude instability of the uniform wavetrains.
We note in passing that
from the linear dispersion relation for the SG equation,
$\omega_{k} = \sqrt{k^{2} +1},$
as $L$ gets larger, more and more discrete
linear phonon modes shift below a fixed $\kappa$, where
the driving is $\sin (\omega t -\kappa x)$. In our case there
is only the $k=0$
phonon mode below $k_{1}$,  2 discrete linear phonon modes  below $k_{2}$,
and 3 below $k_{3}$; the dynamics as $L \rightarrow \infty $, such as the
excitation bifurcation sequence,  will be different from that
depicted here.

 Now we turn to the spatial
instability of the  uniform wavetrains. At the boundary
(Eq(~\ref{eq:critical})), $Z(z)$ has a solution periodic in space and time.
Therefore the
solution of Eq(~\ref{eq:X}) must be periodic in space with \( K^{2} = (2\pi
m/L)^{2} \equiv K_{m}^{2}, m= 1,.... \) The solution $\cos (K_{m} x)$
 is a modulation in space on the amplitude of
$Z(\omega t - k_{n} x)$.
 $K_{1}$ is the lowest spatial modulational mode.
The existence of the mode  $K_{1}$ in the unstable region
 of the uniform wavetrain
 for $\Gamma \geq
\Gamma_{c}$ signifies formation of a new state with a spatial structure
whose length scale may be of order $K_{1}^{-1}$.
$K_{1}^{-1}$ is yet another
length scale in the system near
$\Gamma_{c}$
 in addition to the width of
the excited breather, which is controlled by $\omega$, and the spatial period
of the traveling background $\Gamma \cos(\omega t - k_{n}x)$
on which the breathers ride.
The importance
of this length-scale competition will be further illustrated with the nonlinear
spectral analysis in section V.

\vspace{24pt}
\noindent
IV. ENERGY BALANCE ESTIMATE
\vspace{20pt}

In this section, we point out that a simple energy
balance argument \cite{Mazor2,Lomdahl} can predict the critical value of
$\Gamma_{c}$ in good agreement with the numerical results. The simple
ansatz we use is one breather with a static center, plus a sinusoidal traveling
wave which is the solution for the linearized Eq(~\ref{eq:SG})
(See Eq(~\ref{eqf})):
\begin{eqnarray}
U(x,t) & = & U_{B}(x,t) + f \\
U_{B}(x,t) & = & 4 \arctan [ \frac{(1-\Omega^{2})^{1/2}}{\Omega}
\frac{\cos(\Omega t - \Theta)}{\cosh [(1-\Omega^{2})^{1/2} x]}],
\end{eqnarray}
where $U_{B}(x,t)$ is the pure SG breather with $\Omega = \omega =0.9$. From
the
discussion above, the reason for this choice of $\Omega$ is clear;
assuming the breathers to be traveling at a slow speed without much
interaction,
so that the ``relativistic'' correction to energy is small,
 our ansatz  neglecting the interaction is
reasonable. In section V we will see that the above assumption is valid.

  The change in energy in a period of time $T = 2\pi/\omega$
is
\begin{equation}
\Delta H =\int_{-L/2}^{L/2} dx \int_{0}^{T}dt(\Gamma \sin(\omega t - k_{n} x) -
\alpha U_{t}) U_{t}.  \label{eq:fred}
\end{equation}
If we assume that the energy input to and output from the system
in one period is balanced, {\it i.e.} $\Delta H = 0,$ and
that  $\tilde{\Gamma}_{c}$ is
the smallest $\Gamma$ which
 can balance
the above equation, then from Eq(~\ref{eq:fred}) it
is easy to obtain
\begin{equation}
\tilde{\Gamma}_{c} = \frac{4\alpha \sqrt{1-\omega^{2}}}{\omega}
\frac{1}{I(k_{n})}
\arctan {\frac{\sqrt{1-\omega^{2}}}{\omega \sqrt{1+\frac{1}{\omega^{2}
(\sinh^{-1}[L\sqrt{1-\omega^{2}}/2])^{2}}}}}, \label{eq:gamma}
\end{equation}
where \(I(n)=\int_{-L/2}^{L/2}dx \cos( k_{n}x) (\sqrt
{\cosh^{2}(x \sqrt{1-\omega^{2}}) + (1-\omega^{2})/\omega^{2}} - \cosh (x
\sqrt{1-\omega^{2}}))\).  Comparing estimate (~\ref{eq:gamma}) with
our numerical simulations,  we have the results in Table~\ref{T1}
(See also Fig.~\ref{f1}).

\vspace{24pt}
\noindent
V. NONLINEAR SPECTRAL ANALYSIS
\vspace{20pt}

Numerical nonlinear spectral analysis is a powerful tool for
diagnosing the solitonic content and radiation background for
systems such as
the weakly perturbed  NLS and SG equations. The general theory and
the numerical implementation in these cases
are found in Refs \cite{Overman,Forest,Flesch}.
In the following, we will use this analysis to deepen our understanding
of the behavior of our system and also to show that there is a simple
mechanism which sets the excited breathers into translation.

As mentioned in section II, our system forms a periodic traveling wave locked
to driving from the flat initial background.
 With increasing $\Gamma$
the system exhibits a sharp transition at $\Gamma_{c}(n)$ to a new coherent
state where, for $n = 1, 2,$ a {\em 2-breather} excitation
 is detected by the nonlinear spectral analysis.  For $n =1$,
these two breathers are out of phase and for $n=2$, they are in phase.
Frequencies determined from the spectral analysis show that
it is a good approximation that two breathers are locked to the
temporal frequency $\omega$, see Fig.~\ref{f2}. For $n=2$, the frequencies of
the breathers have a larger deviation from $\omega$. They oscillate
 between 0.75 and 0.9.
  Also the two excited breathers are
observed to run, {\it i.e.} translate along the system,  at an average speed,
$0.25$ for $n=1$, $0.48$ for $n=2$,
with some small fluctuations around these average speeds.  Notice that
the group velocity $V_{g}$ is
\begin{equation}
V_{g} = \frac{k}{\sqrt{k^{2} +1}}\label{eq:vg}
\end{equation}
from the linear dispersion relation for the SG equation:
\begin{equation}
\omega_{k} = \sqrt{k^{2} +1}. \label{eq:dispersion}
\end{equation}
For
$k_{n} = 2 \pi n/L, V_{g}(n=1)= 0.25, V_{g}(n=2)= 0.46$,
respectively. These values coincide with the
average speed of the breathers within our error estimate.
 Since the
value of  $\omega$ is below the threshold of the phonon frequency (see
Eq(~\ref{eq:dispersion})), only the spatial part of the driving
can select the
wavenumber of the phonons. The nonlinear spectrum indeed shows the $k_{n}$
excitation as a radiation spine attached to the real axis: it appears
that these phonons drive the breathers into motion at the group
velocity if the breathers are viewed as a disturbance around those
 $k_{n}$'s. We have  confirmed
this assumption with $L = 36$, See Table~\ref{T2}.
Our system has
a driving force that couples the degrees of freedom in time and space.
Nonetheless the above results indicate  that in the parameter region of
interest the frequency and the velocity of the excited breathers
are controlled separately by the temporal and the spatial parts of
the driving, respectively.

 We also searched for a possible window of 1-breather
excitation which might occur near $\Gamma_{c}$ following the
traveling
wave region and before the onset of the 2-breather excitation. We did
not find any such window for $L=24$. We have examined
$\Delta\Gamma/\Gamma_{c}$ to $10^{-3}$,
where $\Delta\Gamma$ is the width of the
window in which a one-breather excitation might exist.

As $\Gamma$ is
increasing further above $\Gamma_{c}$, the centers of the two breathers
start to move relative to each other, and it appears that they
are weakly interacting. A third breather
can be created if $\Gamma$
 is sufficiently increased. The frequencies of these three
breathers oscillate around $\omega$ and they exhibit
rather complicated structure in time:  These two breathers have frequencies
about 0.9 and 0.7, respectively, and the other
one has very unstable structure  often
collapsing into anharmonic radiation and then re-emerging.
As a typical case, consider
$L = 24, n=1, \Gamma=0.3$: for $0 \leq t <1900$,
there are only two-breather excitations; for $t>1900$, there are 3
breathers. Many features of the system are still describable in terms
of the behavior of these breathers. For example, the modulation
of the total energy in time is discernable as the modulation
due to the different frequencies of these breathers. These
observations reinforce the notion that the collective coordinate
description for this nearly integrable system  captures essential
aspects of the system when a small number of degrees of freedom is involved.

 For $n=3$, the transition is again sharp. The amplitude of the field is
driven  to the order of magnitude $2\pi$. Then, new kink excitations
together with breather excitations are detected by the nonlinear spectral
analysis. The behavior of the  loci of these solitonic spectral
bands nonetheless indicates that the
spatially coherent structure is less salient and is close to
 being destroyed: the field has the form of a traveling
wave packet with a large modulation than any spatially coherent excitation.
It is hard to measure the group velocity of the wave packet in this case
because it constantly alters its shape while traveling.
However, a rough estimate of $V_{g}$ again agrees with the picture
discussed above (see Eq(~\ref{eq:vg})).

To illustrate
the competition of the three length-scales and the importance
of the $K_{1}$ mode in controlling the dynamics, we select the case of $L=36,$
$n=2,$$\Gamma = 0.207$: From $t= 0$ to $t \simeq 160$, the system is locked to
the
external driving as the sinusoidal amplitude of the field  increases.
Around $t > \sim 160$, four
breathers are excited (see Fig.~\ref{f3}(a)). Then
a new state competes with this four-breather state. It has the form of
 a locked traveling wave with a large
deviation away from sinusoid. Nonlinear
spectral analysis shows a long radiation spine
(see Fig.~\ref{f3}(b)). The duration $\Delta t$
of each state is  about 200- 300 time units. These two states
interleave with each other until $t \simeq 920$ after
which the system settles into an attractor
 with two traveling breathers
immersed in a low-amplitude radiation background. These two breathers have a
pointlike spectral band in contrast to the relatively long spectral bands of
the four breathers competing with the locked state (see Fig.~\ref{f3}(c)).

In all cases, $L=24$ and $36$, $n=1,2$, the system is attracted
to a two-breather state at the first bifurcation after the transient if the
length of the system can accommodate two such breathers whose frequencies
are equal to $\omega$ in the lowest order of approximation. It appears that
the amplitude instability with the $\cos(K_1x)$ modulation
leads to the formation of the two-breather state.
We therefore propose that it is the $K_{1}$ mode that is most
important in controlling the dynamics in this parameter region.

\vspace{24pt}
\noindent
VI. CONCLUSION
\vspace{20pt}

We have identified two distinct regimes for our damped sine-Gordon chain
driven by the spatio-temporal periodic potential.
We have demonstrated that in regime I, of the three relevant length
scales of our system, the $K_{1}$ mode appears to be the most important
for controlling the low excitations, and that, in regime II,  $k_{n}$
is the controlling length-scale for the excitations of the system.
 We have  discussed different
selection mechanism for determining the temporal frequencies of the breathers,
 controlled by the temporal part of the driving, and
the running speed of the breathers, controlled by the spatial
part of the driving. It appears that in our parameter region
the presence of the
 traveling background causes the system to tend to form two running breather
excitations at the first bifurcation in contrast to the case of pure temporal
ac-driving where the system is first attracted to a state of one
non-translating breather.
We can expect, of course, that breathers excited
by an ac-{\em standing-wave}-like driving will not run.

In summary,
this work provides a further example supporting the concept of collective
coordinates and competing length scales
for soliton systems under relatively weak perturbation.

\vspace{24pt}
\noindent
ACKNOWLEDGMENTS
\vspace{18pt}

We thank E.A. Overman for his nonlinear spectral code
and his advice on the numerical IST, and R. Scharf and P. Lomdahl for
discussions. We acknowledge support for this work by the U.S. D.o.E. and
partial financial support for A.S. by CICyT (Spain) under project MAT90-0544.

\pagebreak

\begin{table}
\caption{Comparison of the measured $\Gamma_{c}^{measure}$ with
$\tilde{\Gamma}_{c}$ estimated by the energy balance argument and
$\Gamma_{c}$ estimated from linear stability analysis (see text).}
\smallskip

\begin{tabular}{||c|c|c|c||}
$n$		& 1		&2		&$3$		\\ \hline
$\Gamma_{c}^{measure}$ &0.1643-0.1645 &0.3421-0.3425	&~~~~0.7560~~~~~\\ \hline
$\tilde{\Gamma}_{c}$(fr. Eq(~\ref{eq:gamma}))& 0.144	&0.345	&0.891	\\ \hline
$\Gamma_{c}$ &0.24  &0.53  &1.19
\end{tabular}
\label{T1}
\end{table}

\begin{table}
\caption{$V_{g}^{measure}$ is the measured
average velocity of a breather. $V_{g}$
is the group velocity  calculated from Eq(~\ref{eq:vg}).}
\smallskip

\begin{tabular}{||c|l|l|l|l||}
$L$		&24	&24	&36	&36	\\ \hline
$n$		&1	&2 	&1	&2	\\ \hline
$V_{g}^{measure}$ &0.25	&0.48	&0.18	&0.34	\\ \hline
$V_{g}$ 	&0.25	&0.46	&0.17	&0.33
\end{tabular}
\label{T2}
\end{table}

\figure{\label{f1}
Bifurcation diagram for the damped sine-Gordon chain driven by
a spatio-temporal periodic force in the NLS regime. Parameters:
$\omega=0.9$, $\alpha=0.1$, $L=24$; The crosses are the results
obtained from the numerics, the squares from
the linear stability analysis, and the triangles
from the energy balance estimate.}
\figure{\label{f2}
(a) The field snapshots and the nonlinear spectral bands of the breathers
at $t=4000$ for $n=1$, $\Gamma=0.164534375$, and $L=24$.
The solid line in the inset is $U(x,t)$, and the dashed line is $U_{t}(x,t)$.
(b) The field snapshots and the associated
 nonlinear spectral bands of the breathers
at $t=1988$ for $n=2$, $\Gamma=0.3425781$, and $L=24$.
}
\figure{\label{f3}
The competition of excitations for $\Gamma=.207$, $n=2$, and $L=36$:
(a) a four-breather excitation at $t=740$;
(b) a locked state at $t=770$;
(c) the final state at $t=1760$ after transients.
}

\end{narrowtext}
\end{document}